# Binary Classifier Wire-Resistance Attack on KLJN: Impact of Narrowing the Resistor Gap

MEHMET YILDIRIM[§], LASZLO B. KISH

[1]*Department of Electrical and Computer Engineering, Texas A&M University,*
*College Station, TX 77841-3128, USA*
*eem.mehmetyildirim@tamu.edu, laszlokish@tamu.edu*

**Abstract:** It is shown that narrowing the difference between the high and low resistor values in the Kirchhoff Law-Johnson Noise (KLJN) key exchange strongly affects security against a recently introduced binary classifier-based wire resistance attack. Using time domain simulations of a non-ideal KLJN loop with finite cable resistance, we generate large ensembles of secure (HL/LH) bits and evaluate the mean-square noise voltages at Alice's and Bob's ends. For each bit, these mean-square values form a point in a two-dimensional classifier plane, where the separation between the HL and LH point clouds characterizes the information available to an eavesdropper (Eve). We quantify Eve's success probability *p* by a simple decision rule based on the sign of the difference between the measured mean-square voltages. For strongly asymmetric resistors (for example $R_L$ = 4 kOhm and $R_H$ = 10 kOhm) and realistic wire resistances, the HL and LH clouds are fully separable and Eve's *p* approaches 1, which confirms that the classifier attack can practically recover all secure bits. As the low resistor value approaches the high one (for example $R_L$ = 9 kOhm and $R_H$ = 10 kOhm) at the same cable resistance, the HL and LH clouds increasingly overlap and the measured *p* drops close to 0.7, approaching the ideal limit *p* = 0.5 as $R_L$ approaches $R_H$. A surprising phenomenon is that, in this classifier-based scenario, increasing the wire resistance can decrease the information leak. This counterintuitive effect is strikingly the opposite of the behavior in the classical Bergou–Scheuer–Yariv wire resistance attack, where the mean-square voltages at the two ends of the wire are simply compared.

## 1. Introduction

Information-theoretic (unconditional) security [1-4] aims to guarantee that a cryptographic system remains secure even against an adversary with unlimited computational power, so that the observed signals reveal no useful information about the secret data. In any such scheme, the security of the overall communication cannot exceed the security of its key exchange protocol. So far, only two hardware-based key exchange mechanisms have been proposed that can claim information-theoretic security: quantum key distribution (QKD) [5-43] and the classical Kirchhoff Law-Johnson Noise (KLJN) key exchange [3,4, 44-105]. The latter (Figure 1) relies solely on classical statistical physics, using thermal (Johnson)

[§] Corresponding Author

noise and Kirchhoff's circuit laws, and has been extensively analyzed over the past two decades.

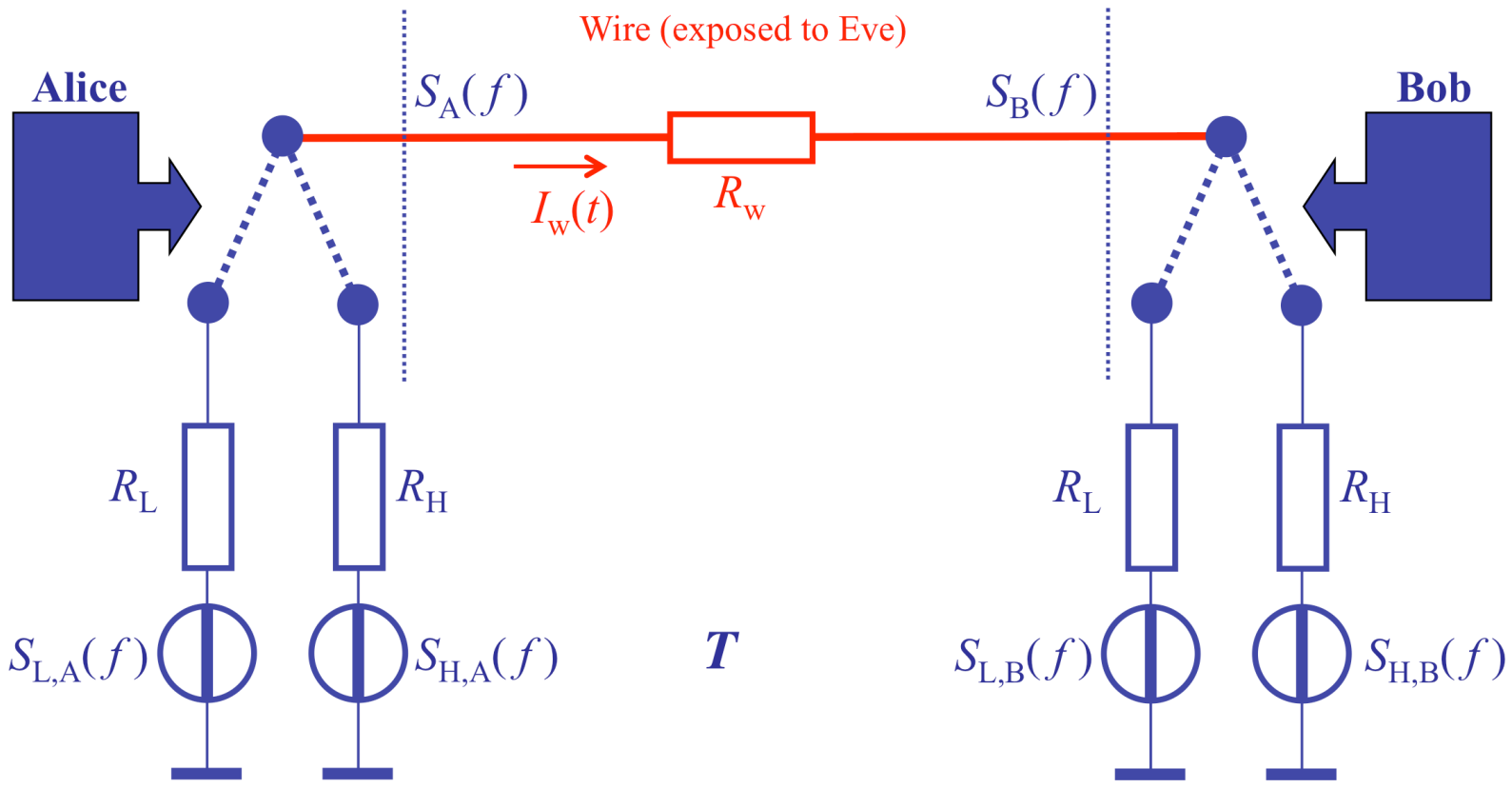

Fig. 1. The KLJN scheme with non-zero wire (cable) line resistance $R_w$ [44-49]. $T$ is the common (effective) temperature, the four independent noise voltage generators are the thermal noises of the resistors (they are typically emulated by active generators for enhanced temperature); $U_{L,A}(t)$, $U_{L,B}(t)$, $U_{H,A}(t)$ and $U_{H,B}(t)$ are their Johnson noise voltage. $U_A(t)$ and $U_B(t)$ are the resultant noise voltages at Alice's and Bob's end on the wire line and $I_w(t)$ is the current in it.

In the ideal KLJN scheme [44], Alice and Bob are connected by a perfect zero-resistance wire, and each side randomly selects one of two resistors, denoted RL and RH, representing the low and high bit values. During a given clock period, the combination where Alice and Bob choose different resistors (HL or LH) constitutes a secure bit situation. In this ideal limit, the mean-square noise voltage and current on the wire are identical for the HL and LH cases, and a passive eavesdropper (Eve), who can only measure voltage and current on the line, cannot distinguish which side holds which resistor. Thus, in the absence of nonidealities and active interference, the mixed resistor states are information-theoretically secure.

However, practical KLJN implementations inevitably deviate from this idealized model. Nonzero wire resistance, cable capacitance, finite temperature accuracy, transients, and other imperfections break the perfect symmetry between the HL and LH states and open the door to various attacks. One of the earliest and most studied is the wire resistance (Bergou–Scheuer–Yariv type) attack [45-47], where the finite resistance of the cable causes small but systematic differences between the noise statistics at Alice's and Bob's ends.

More recently, a binary-classifier-based [104] wire-resistance attack that we have nicknamed "Mehmet-attack" has been introduced. In this attack, Eve records the mean-square noise voltages at both ends of the line over many secure bits and represents each bit by a point in a two-dimensional plane defined by these mean-square values. For realistic, nonzero wire resistance, the HL and LH bits form two well separated point clouds, enabling Eve to classify the bits with success probability $p$ close to 1, without requiring sophisticated machine learning tools.

So far, most countermeasures against such wire-resistance-based attacks have focused on compensating the nonidealities (for example by adjusting the effective noise temperatures at the two ends) [52,89], on discarding high risk bits [63,65,66], or on applying privacy amplification to the raw key [62].

Another, more recently explored approach [105] is to reduce Eve's information already at the physics level by modifying the KLJN parameter values. In particular, earlier work has shown that narrowing the gap between the resistor values RL and RH increases Alice's and Bob's bit error probability at a fixed clock period, but at the same time it reduces the information leak in various asymmetry-based passive and active attacks. This trade–off suggests that making the resistor pair more symmetric may "make the eavesdropper's life harder", at the price of lower communication speed.

In the present paper, we investigate how this resistor-narrowing strategy affects the strength of the binary-classifier-based wire-resistance attack and, more generally, what this implies for integrated circuit hardware security applications of KLJN. Using time domain simulations of a non-ideal KLJN loop with finite cable resistance, we generate large ensembles of secure bits and evaluate the mean-square noise voltages at Alice's and Bob's ends. Each secure bit is mapped to a point in the classifier plane, and we quantify Eve's success probability $p$ by a simple decision rule based on the sign of the difference between the measured mean-square voltages. By systematically varying the low resistor value from strongly asymmetric choices up to nearly the high resistor value while keeping the wire resistance fixed, we show how the separation between the HL and LH point clouds shrinks and how Eve's $p$ moves from practically 1 toward 0.5. A further surprising result is that, in this classifier-based scenario, increasing the wire resistance can actually reduce the information leak, which is strikingly opposite to the trend observed in the classical Bergou–Scheuer–Yariv [47] wire resistance attack, where only the mean-square voltages at the two ends of the wire are compared. These findings provide quantitative guidance on how resistor narrowing, together with realistic interconnect resistance, can be used at the circuit level to reduce side-channel leakage and clarify the trade-off between security and key exchange speed in potential on-chip and chip-to-chip KLJN implementations aimed at unconditionally secure hardware communication.

## 2. Simulations of narrowing the resistance gap

In preliminary simulations, generating the noise at Alice and Bob from disjoint segments of the same random-number stream led to a small but systematic non-zero average power flow between the two sides in the secure states. This behavior contradicts the zero-power-flow condition in ideal KLJN and indicates that the two noise sources were not statistically independent in the simulation. When we instead used two independent Gaussian random streams initialized with different seeds, the average power flow vanished, restoring the expected KLJN behavior.

In Figures 2-5, the results of the Mehmet attack are shown for resistances 10-10 kOhm, 9-10 kOhm, 7-10 kOhm, and 4-10 kOhm at 200 Ohm (Figures a) and 800 Ohm (Figures b) wire resistances, respectively.

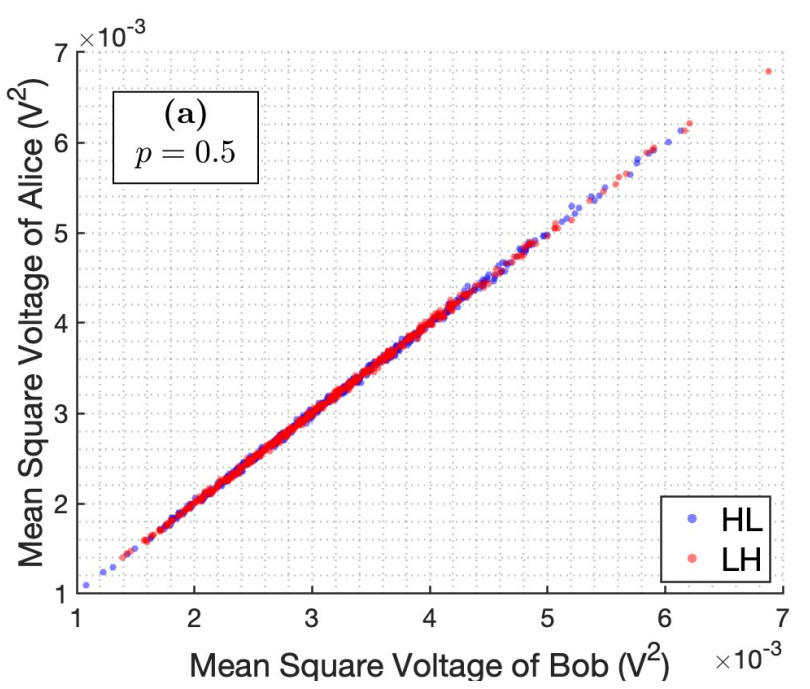

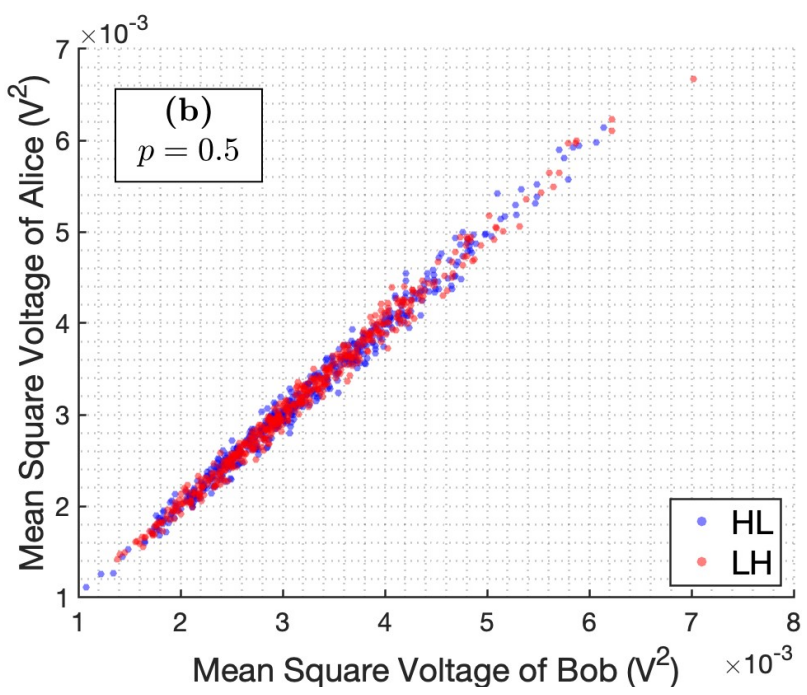

Fig. 2. Mehmet-attack plot of the mean-square voltages of Alice and Bob for uniform resistances $R_L$ and $R_H$ (both 10 kOhm) with (a) 200 Ohm and (b) 800 Ohm wire resistance. Eve's successful bit guessing probability is $p = 0.5$ because with equal resistances she cannot extract any information, similarly to Alice and Bob.

Figure 2, with $R_L$ and $R_H$ both equal to 10 kOhm (thus no key exchange is possible $p = 0.5$), shows how $R_w$ decouples the fluctuations at Alice's and Bob's sides. At $R_w = 800$ Ohm, a much larger number of dots scatter away from the immediate vicinity of the $x = y$ line.

Figure 3 shows that, due to the different resistances of Alice and Bob, the dots now scatter around lines with different slopes. This means that Alice and Bob can perform key exchange, but part of this information also leaks to Eve. The resulting values of $p = 0.706$ and $p = 0.704$ further indicate this leakage.

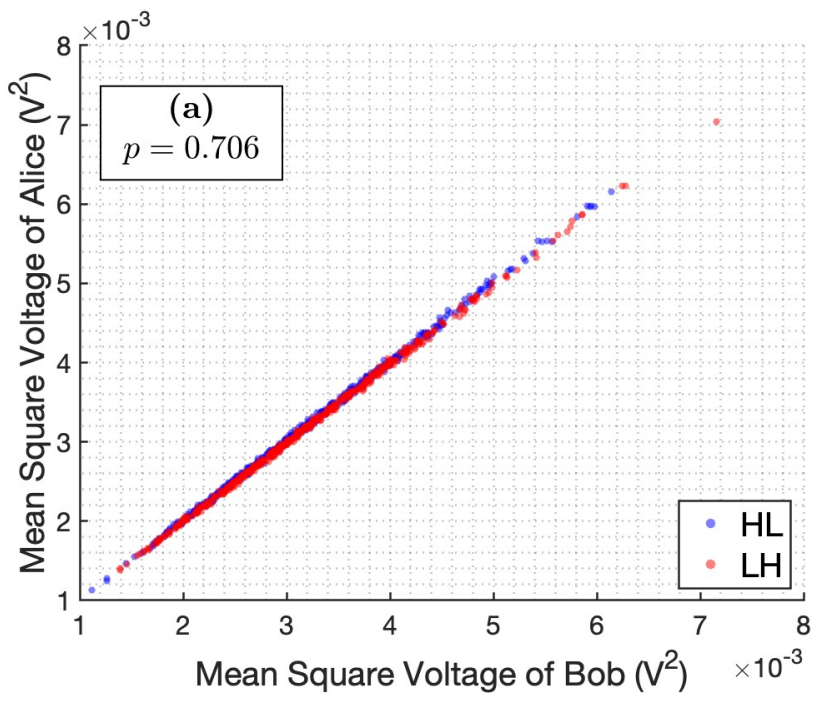

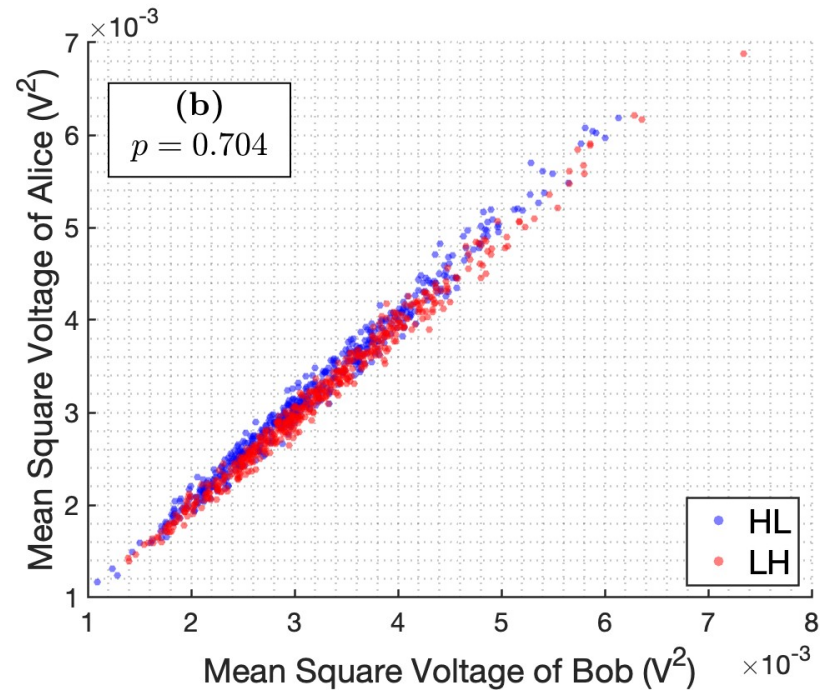

Fig. 3. Mehmet-attack plot of the mean-square voltages of Alice and Bob for $R_L$=9 kOhm and $R_H$=10 kOhm with (a) 200 Ohm and (b) 800 Ohm wire resistances. Eve's successful bit guessing probabilities are $p$ = 0.706 and $p$ = 0.704, respectively.

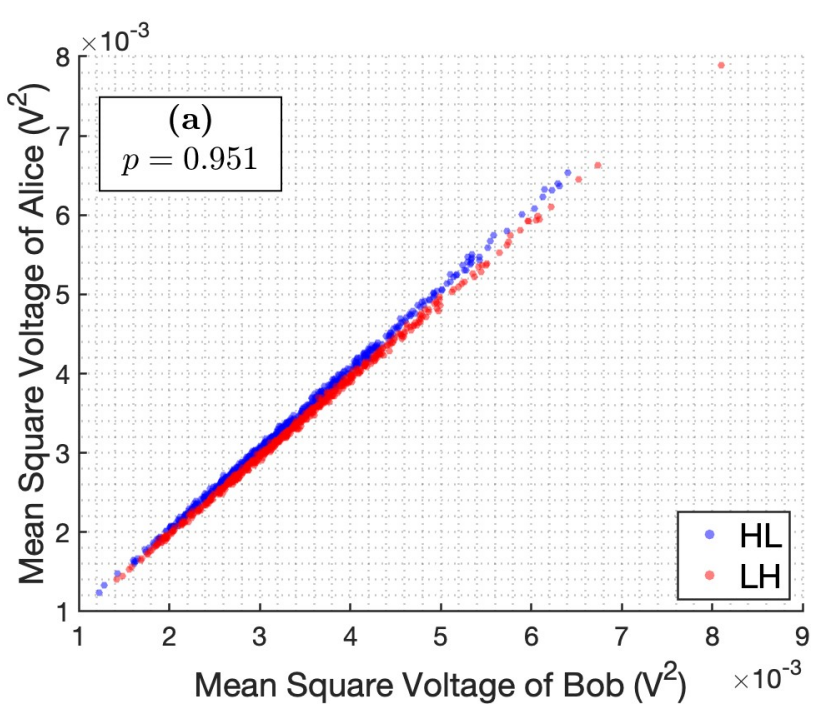

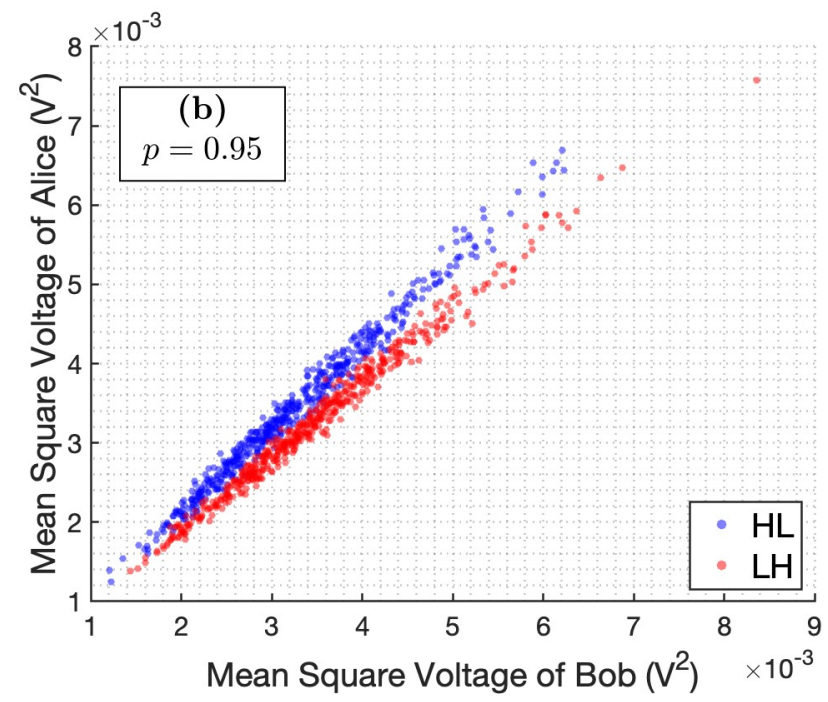

Fig. 4. Mehmet-attack plot of the mean-square voltages of Alice and Bob for $R_L$=7 kOhm and $R_H$=10 kOhm with (a) 200 Ohm and (b) 800 Ohm wire resistances. Eve's successful bit guessing probabilities are $p$ = 0.951 and $p$ = 0.95, respectively.

Figure 4 shows a similar trend: the increasing resistance gap (10 kOhm, 7 kOhm) promotes the efficiency of the Mehmet attack. The resulting values of $p$ = 0.951 and $p$ = 0.95 further indicate this leakage.

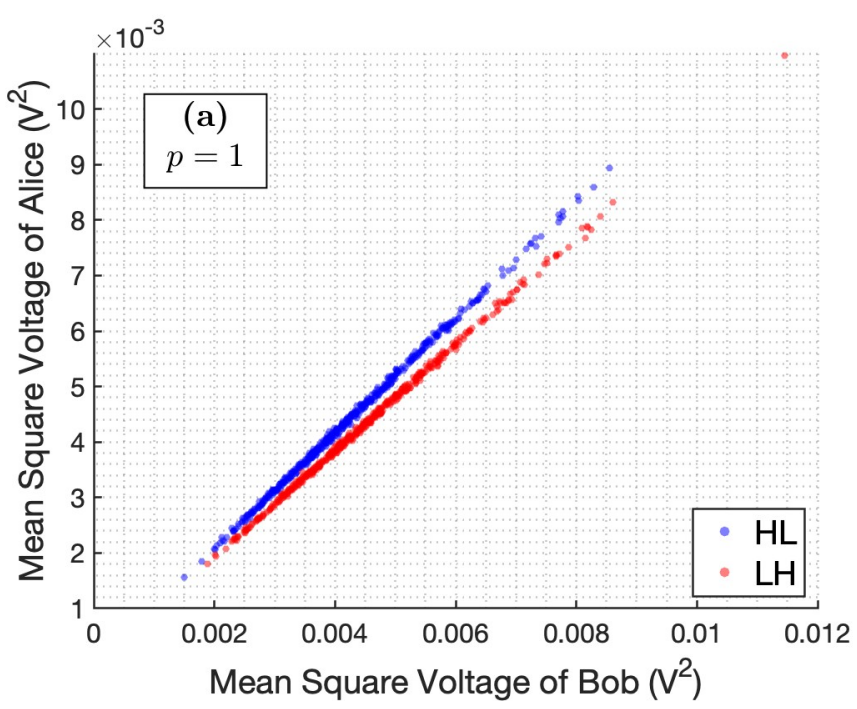

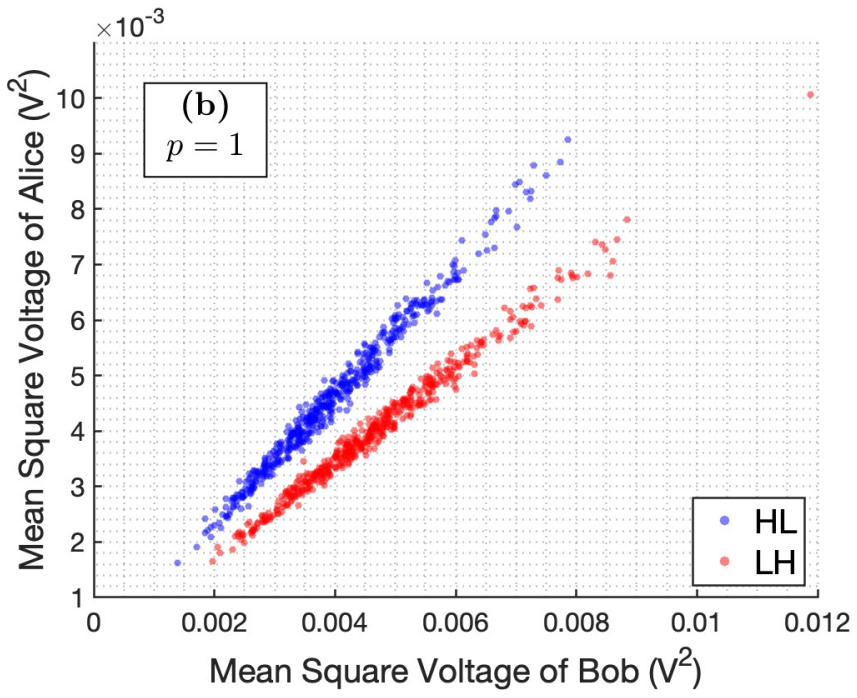

Fig. 5. Mehmet-attack plot of the mean-square voltages of Alice and Bob for $R_L$=4 kOhm and $R_H$=10 kOhm with (a) 200 Ohm and (b) 800 Ohm wire resistances. Eve's successful bit guessing probabilities are $p$ = 1 and $p$ = 1, respectively.

The same trend continues in Figure 5 for $R_L$=4 kOhm and $R_H$=10 kOhm.

## 3. Surprising impact of the wire resistance

Figure 6 shows the explicit information leak by plotting Eve's probability $p$ of successfully guessing the actual bit values over a range of wire resistances. At the first glance, the data are surprising because $p$ decreases monotonically with increasing wire resistance. This indicates that the Mehmet attack realizes less information leak at higher wire resistances, which is opposite to the behavior of the former attacks based on comparing the mean-square voltages at the two ends of the wire channel.

The interpretation of this counterintuitive effect is straightforward: with increasing wire resistance, the correlations between the noise voltages at the two ends of the wire are decreasing because the wire resistance tends to decouple the noise generators of Alice and Bob. Thus, the efficiency of the Mehmet attack falls, while the efficiency of the classical, mean-square-comparison-based attack improves as Eve's signal scales with the square of wire resistance (in the low-resistance limit).

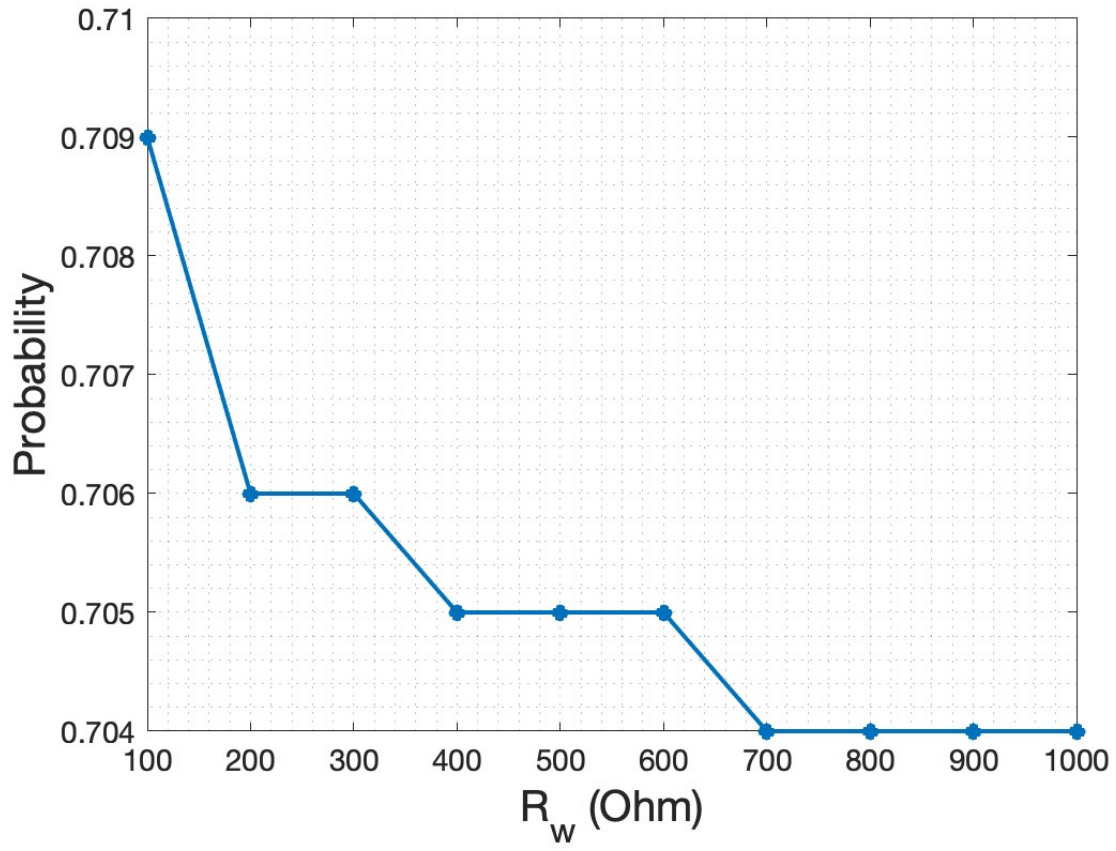

Fig. 6. The $p$ values, at Alice's/Bob's resistors being 9/10 kOhm, versus the wire resistance $R_w$. These results further confirm the counterintuitive nature of the behavior of $p$ versus $R_w$ during the Mehmet attack.

## 4. Conclusions

We have shown that narrowing the resistor gap $R_L \rightarrow R_H$ drives Eve's bit-guessing probability $p$ in the binary classifier (Mehmet) attack from near unity toward the secure limit of 0.5, confirming that resistor symmetrization is an effective physics-level countermeasure. A counterintuitive but key finding is that increasing the wire resistance $R_w$ reduces the information leak in the classifier attack — the opposite of the classical Bergou–Scheuer–Yariv attack, where $p$ grows rapidly with $R_w$ because Eve's signal scales with the square of wire resistance [47].

This opposing behavior suggests a natural and practical optimization criterion: since $p$ falls only slowly with $R_w$ under the classifier attack while rising rapidly under the classical attack, there therefore exists a crossover wire resistance at which both attacks yield similar $p$. Operating near this crossover point minimizes the maximum advantage available to Eve across both attack strategies simultaneously, providing a "minimax" security optimum.

Importantly, this optimization has a direct engineering benefit: higher wire resistance corresponds to either a longer communication range or the use of thinner wire with less copper, reducing material cost and weight. Together, resistor-gap narrowing and crossover-optimized wire resistance offer a practical, hardware-level design strategy for suppressing side-channel leakage in on-chip and chip-to-chip KLJN implementations aimed at unconditionally secure communication.

A quantitative crossover optimization, based on the theoretical expression for $p$ in the classical Bergou–Scheuer–Yariv attack, will be the subject of a forthcoming study.